\documentclass{iau}

\usepackage{natbib}
\usepackage{graphicx}
\usepackage[breaklinks,colorlinks,
   urlcolor=blue,citecolor=blue,linkcolor=blue]{hyperref}
\usepackage{lipsum}
\usepackage{ragged2e}
\usepackage{setspace}

\usepackage{multicol}

\usepackage[space]{grffile}
\usepackage{latexsym}
\usepackage{amsfonts,amsmath,amssymb}
\usepackage{url}
\usepackage[utf8]{inputenc}
\usepackage{fancyref}
\usepackage{textcomp}
\usepackage{longtable}
\usepackage{multirow,booktabs}
\usepackage{subfigure}
\usepackage{color}
\usepackage{gensymb}
\usepackage{enumitem}

\newcommand{\zsol}{Z$_\odot$}

\newcommand{\OVI}{[\hbox{{\rm O}\kern 0.1em{\sc vi}}]}
\newcommand{\Lalpha}{Ly$\alpha$}
\newcommand{\NV}{[\hbox{{\rm N}\kern 0.1em{\sc v}}]}
\newcommand{\CII}{[\hbox{{\rm C}\kern 0.1em{\sc ii}}]}
\newcommand{\SiIV}{[\hbox{{\rm Si}\kern 0.1em{\sc iv}}]}
\newcommand{\OIV}{[\hbox{{\rm O}\kern 0.1em{\sc iv}}]}
\newcommand{\NIV}{[\hbox{{\rm N}\kern 0.1em{\sc iv}}]}
\newcommand{\CIV}{\hbox{{\rm C}\kern 0.1em{\sc iv}}}
\newcommand{\HeII}{\hbox{{\rm He}\kern 0.1em{\sc ii}\kern 0.1em{$\lambda1640$} }}
\newcommand{\OIII}{[\hbox{{\rm O}\kern 0.1em{\sc iii}}]}
\newcommand{\NIII}{[\hbox{{\rm N}\kern 0.1em{\sc iii}}]}
\newcommand{\AlIII}{\hbox{{\rm Al}\kern 0.1em{\sc iii}}}
\newcommand{\SiIII}{\hbox{{\rm Si}\kern 0.1em{\sc iii}}]}
\newcommand{\CIII}{\hbox{{\rm C}\kern 0.1em{\sc iii}]}}
\newcommand{\NeIV}{[\hbox{{\rm Ne}\kern 0.1em{\sc iv}}]}
\newcommand{\MgII}{\hbox{{\rm Mg}\kern 0.1em{\sc ii}}}

\newcommand{\HeIIoptical}{\hbox{{\rm He}\kern 0.1em{\sc ii}\kern 0.1em{$\lambda4686$} }}
\newcommand{\He}{\hbox{{\rm He}\kern 0.1em{\sc ii}}}
\newcommand{\CIIIL}{\hbox{{\rm C}\kern 0.1em{\sc iii}]\kern 0.1em{$\lambda1907,\lambda1909$}}}

\newcommand{\Hep}{He$^+$}

\newcommand{\SII}{[\hbox{{\rm S}\kern 0.1em{\sc ii}}]}
\newcommand{\NII}{[\hbox{{\rm N}\kern 0.1em{\sc ii}}]}
\newcommand{\OII}{[\hbox{{\rm O}\kern 0.1em{\sc ii}}]}

\newcommand{\MgI}{\hbox{{\rm Mg}\kern 0.1em{\sc i}}}
\newcommand{\FeII}{\hbox{{\rm Fe}\kern 0.1em{\sc ii}}}

\newcommand{\OI}{\hbox{{\rm O}\kern 0.1em{\sc i}}}
\newcommand{\NeII}{[\hbox{{\rm Ne}\kern 0.1em{\sc ii}}] }
\newcommand{\NaI}{[\hbox{{\rm Na}\kern 0.1em{\sc i}}] }
\newcommand{\NeIII}{[\hbox{{\rm Ne}\kern 0.1em{\sc iii}}] }

\title[MUSE \HeII\ analysis at $z=2-4$] %% give here short title %%
{Exploring \HeII\ emission line properties at $z=2-4$}

\author[Themiya Nanayakkara, Jarle Brinchmann \& The MUSE Collaboration]   %% give here short author list %%
{Themiya Nanayakkara$^1$,
Jarle Brinchmann$^{1,2}$,
 \and The MUSE Collaboration}

\affiliation{$^1$Leiden Observatory, Leiden University, PO Box 9513, 2300 RA Leiden, The Netherlands \\ email: {\tt nanayakkara@strw.leidenuniv.nl} \\[\affilskip]
$^2$Instituto de Astrofisica e Ciencias do Espaco, Universidade do Porto, CAUP,\\ Rua das Estrelas, 4150-762 Porto, Portugal. \\email: {\tt jarle@strw.leidenuniv.nl}}

\pubyear{2018}
\volume{347}  %% insert here IAU Symposium No.
\jname{Early Science with ELTs}
\editors{Norbert Przybilla, Karen Pollard \& Annalisa Calamida}
\begin{document}

\maketitle

\begin{abstract}
\He\ is the most sought-after emission line to detect and characterize metal free stellar populations. 
However, current stellar population/photo-ionization models lack sufficient \Hep\ ionising photons to reproduce observed \He\ fluxes while being consistent with other emission lines. 
Using $\sim10-30$ hour deep pointings from MUSE, we obtain $\sim10$ $z\sim2-4$ \HeII\ emitters to study their inter-stellar medium  and stellar population properties. 
Emission line ratio diagnostics of our sample suggest that emission lines are driven by star-formation in solar to moderately sub-solar ($\sim 1/20$th) metallicity conditions. 
However, we find that even after considering effects from binary stars, we are unable to reproduce the \HeII\ equivalent widths. 
Our analysis suggest that extremely sub-solar metallicities ($\sim1/200$th) are required to reproduce observed \HeII\ luminosities. Thus, current stellar populations may require alternative mechanisms such as sub-dominant active galactic nuclei  or top heavy initial-mass-functions to compensate for the missing \Hep\ ionising photons.   
\end{abstract}

\firstsection % if your document starts with a section,
\section{Introduction}

The detection and characterization of the first generation of stellar populations in the Universe is of highest priority to the high redshift galaxy evolution community. 
Multiple observational attempts have been made to observationally confirm galaxies with evidence for population III (pop III; metal free) stars without any success  \cite[e.g.,][]{Cassata2013,Sobral2015}, where  \Lalpha\ and \He\ in the absence of other prominent emission lines are interpreted as existence of pristine metal-poor stellar populations \citep[e.g.,][]{Inoue2011,Sobral2015}. 
This interpretation is however challenging in the face of other processes that can produce \Hep\ ionising photons (E$> 54.4$ eV, $\lambda<228$ \AA). 
Additionally, the short life-time of pop-III systems and resulting inter-stellar medium (ISM)/inter-galactic-medium pollution by pair-instability supernovae \citep{Heger2002}, uncertainties in photometric calibrations, presence of active galactic nuclei (AGN), pristine cold mode gas accretion to galaxies, limited understanding of high-redshift stellar populations and the ISM contribute further to the complexity of detecting and identifying pop-III host systems \citep[e.g.,][]{Matthee2017,Shibuya2017,Sobral2017}. 
Thus, to make compelling constraints of stellar populations in the presence of strong \He\ emission and link with pop-III hosts, a comprehensive understanding of \He\ emission mechanisms is required.

Multiple mechanisms prominent in stellar populations in a variety of ages and physical/chemical conditions are expected to contribute to \He\ emission, i.e. young O/B type stars \citep[e.g.,][]{Shirazi2012}, hydrogen-stripped massive evolved Wolf-Rayet stars \citep[e.g.,][]{Shirazi2012}, post-asymptomatic giant branch stars \citep[e.g.,][]{Binette1994}, X-ray binary stars \citep[e.g.,][]{Casares2017}, radiative shocks \citep[e.g.,][]{Izotov2012}, AGN \citep[e.g.,][]{Shirazi2012} have all been suggested as possible contributers. Additionally, mechanisms such as binary interactions and stellar rotation are expected to prolong the lifetime of young O/B stars extending the total amount of \Hep\ photons present at a given star-formation history \citep[e.g.,][]{Eldridge2017,Gotberg2017}.        
Even with a variety of such mechanisms, present stellar-population/photo-ionization models lack sufficient high-energy photons to produce observed \HeII\ line profiles consistently with other rest-UV emission lines in local and high-$z$ galaxies \citep[e.g.,][]{Shirazi2012,Senchyna2017,Berg2018}.

\section{Data \& Analysis}

The advancement of state-of-the-art sensitive multiplexed optical instruments in 8-10m class telescopes such as the The Multi Unit Spectroscopic Explorer \citep[MUSE;][]{Bacon2010} has allowed us to obtain spatially-resolved spectroscopy of galaxies in this epoch in unprecedented numbers \citep[e.g.,][]{Inami2017}.  
Here, we present an analysis done using deep $\sim10-30$ hour pointings from MUSE obtained as a part of multiple MUSE guaranteed time observation programs \citep[][]{Bacon2015,Bacon2017,Epinat2018,Marino2018}. 
Our observations target \HeII\ emitters at $z=1.93-4.67$.
The Universe at $z\sim2-4$ was reaching the peak of the cosmic star-formation rate density \citep{Madau2014}, where systems were highly star-forming and evolving rapidly giving rise to a diverse range of physical and chemical properties \citep[e.g.,][]{Kacprzak2016,Kewley2016,Steidel2016,Nanayakkara2017,Strom2017}.  
Thus, with MUSE we are able to obtain rest-UV spectroscopy of young, low-metallicity, highly star-forming systems which may give rise to a diverse range of exotic phenomena capable of producing high-energy ionizing photons.

\begin{figure}
\includegraphics[trim = 10 10 10 0, clip, scale=0.70]{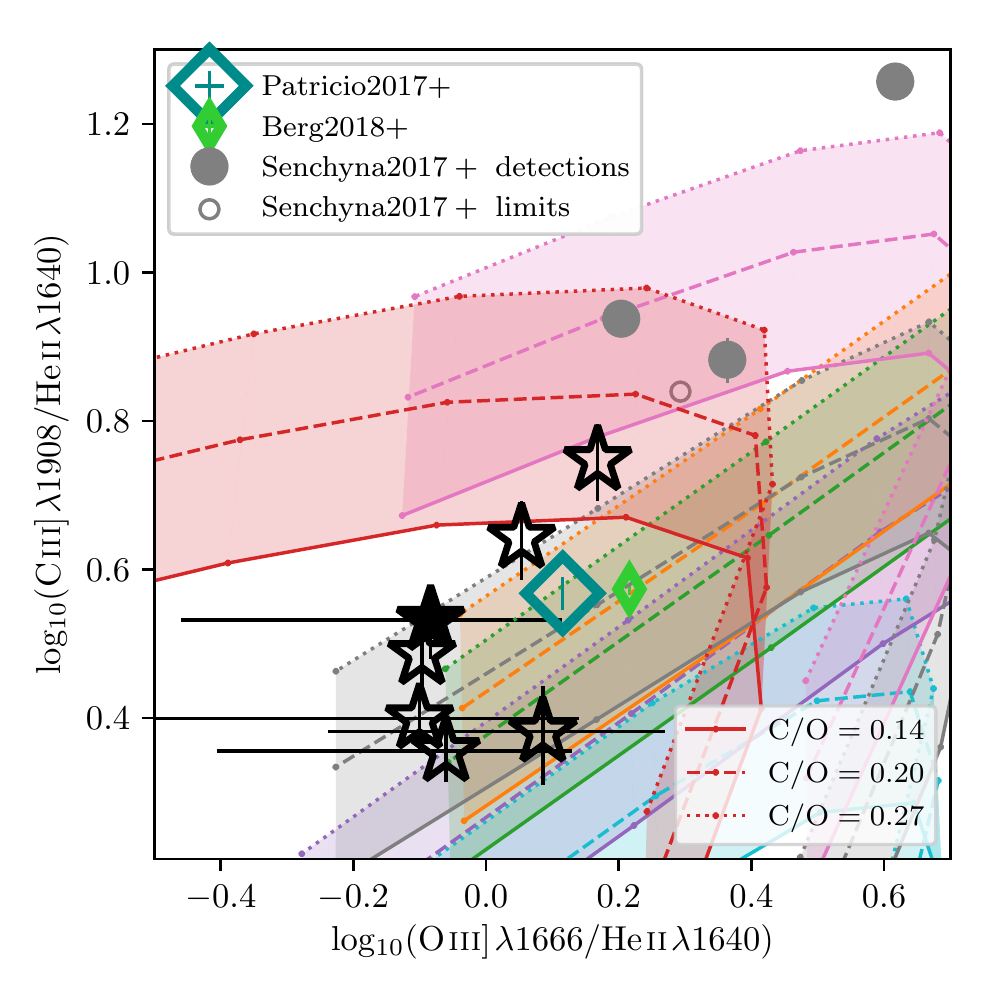}
\includegraphics[trim = 10 10 10 0, clip, scale=0.70]{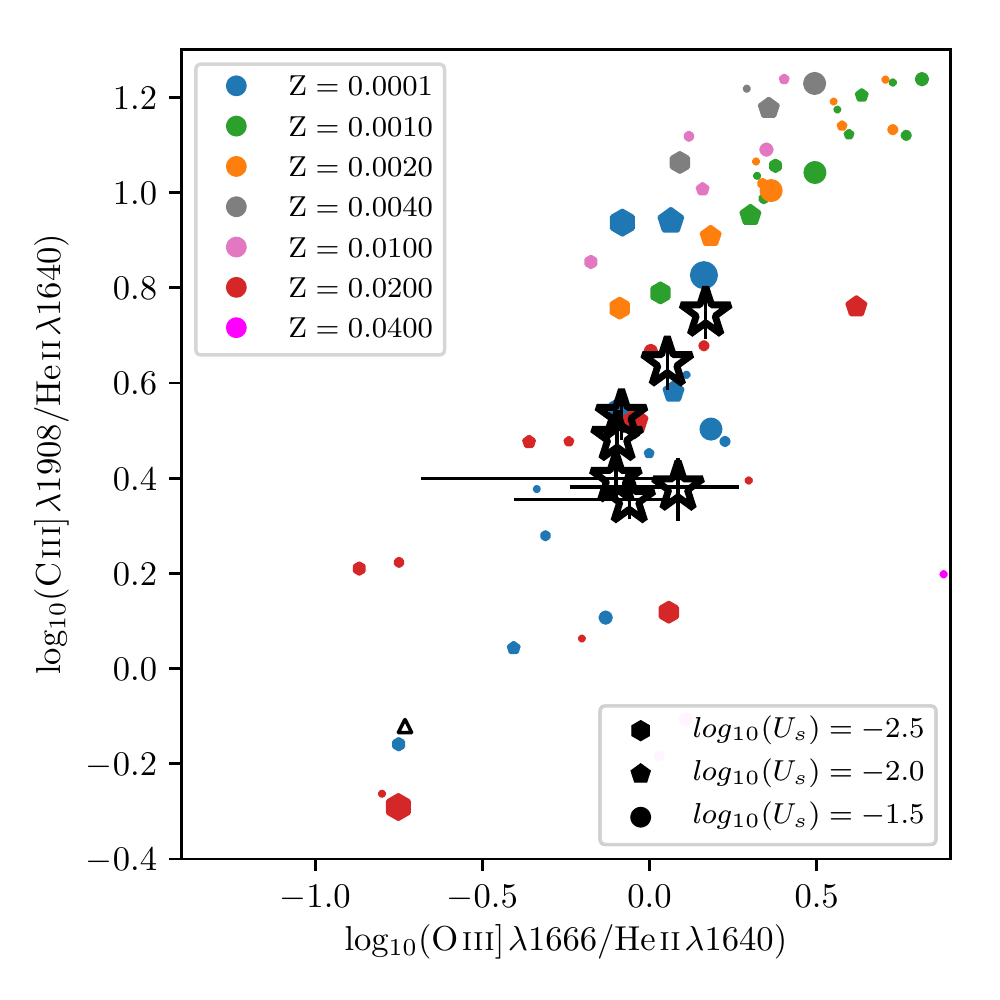}
\caption{Rest-frame  \CIII/\HeII vs \OIII/\HeII\ emission line ratios of the MUSE \HeII\ sample. Individual galaxies with SNR$>3$ for all four emission lines are shown as stars. 
{\bf Left:}  The tracks are from \citet{Gutkin2016} models. Each set of tracks with similar colour resemble three C/O ratios and the region between the minimum and maximum C/O tracks are shaded by the same colour. From top to bottom the ionization parameter increases. 
Line ratios from \citet{Patricio2017}, \citet{Senchyna2017}, and \citet{Berg2018} are shown for comparison. MUSE line ratios of the Lyman continuum emitter from \cite{Naidu2017} is shown by the filled star.
{\bf Centre:} Model line ratios computed by \citet{Xiao2018} using BPASS binary stellar population models.  
Symbol size of each track increase as a function of time and ranges from $t=1$ Myr to $t=50$ Myr from the onset of the star-burst. Models are computed with $log_{10}(n_H)=1.0$ with varying $U_s$ between --2.5 and --1.5. 
\label{fig:line_ratios}
}
\end{figure}

Our sample comprise of 15 \HeII\ detections (including 3 AGN) and is the largest sample of $z>2$ \HeII\ emitters with multiple emission line detections. Additional details on sample selection process is described in Nanayakkara et al., (in prep). 
We remove AGN from our sample and use multiple emission line diagnostics from \citet{Gutkin2016} and \citet{Xiao2018} to explore the ISM conditions of the \HeII\ sample. We find that in  \CIII/\OIII\ vs \SiIII/\CIII, \CIII/\HeII\ vs \OIII/\HeII, and \OIII/\HeII\ vs \CIII/\SiIII\ line ratio diagrams our galaxies occupy a region, that can be described by star-forming galaxies with solar to $\sim1/20$th solar metallicities. 
In Figure  \ref{fig:line_ratios}, we show the \CIII/\HeII\ vs \OIII/\HeII\ line ratio diagrams for single-star stellar population models from \citet{Gutkin2016} and binary-star models from BPASS \citet{Xiao2018}. Our values agree with literature data of high-$z$ sources \citep{Patricio2017,Berg2018} and have considerably lower metallicities compared to $z=0$ sources from \citet{Senchyna2017}. 
When effects of binary stellar models are added, the line-ratio diagnostics become more degenerate \citep[also see ][]{Xiao2018}, however, line-ratios are still within the range powered by star-formation. 

\begin{figure}
\centering
\includegraphics[trim = 10 10 10 0, clip, scale=0.75]{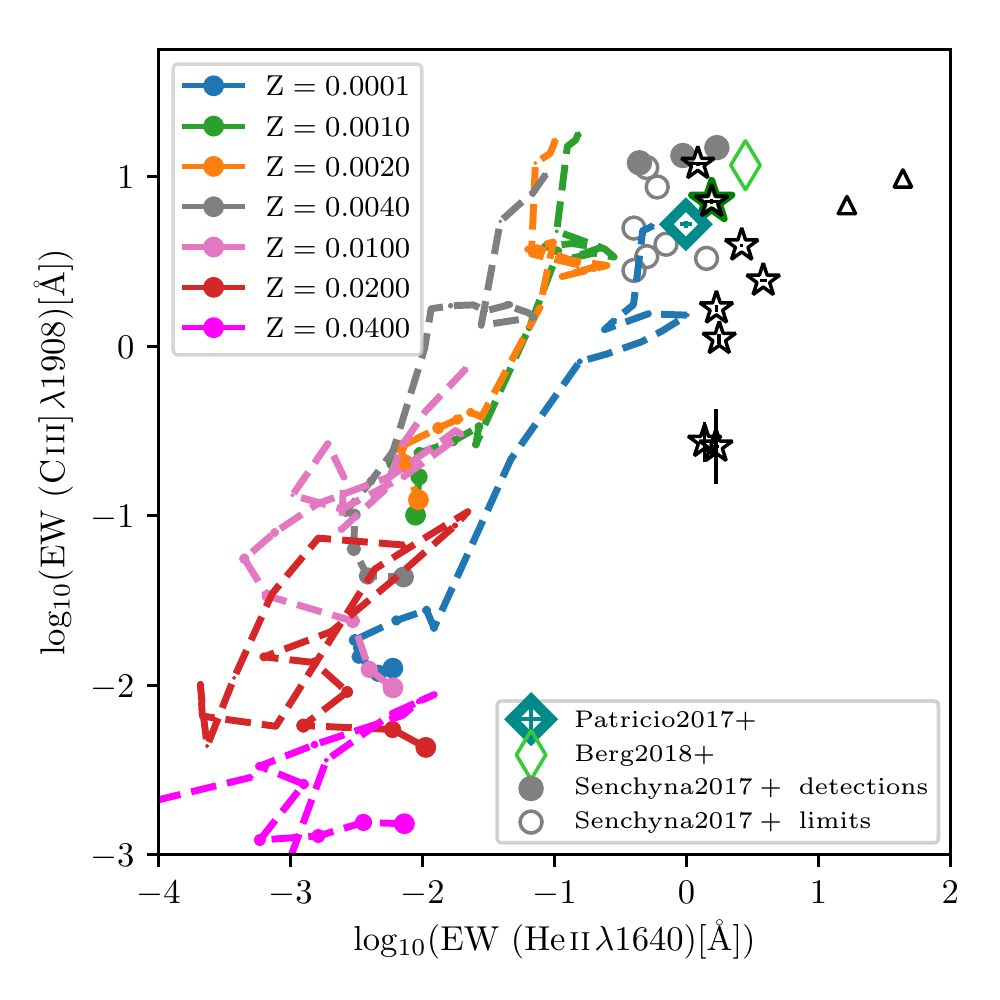}
\caption{\CIII\ vs \HeII\ equivalent widths of \citet{Xiao2018} models for a star-burst stellar population. Model parameters are similar to Figure \ref{fig:line_ratios} right panel. 
\label{fig:ews}
}
\end{figure}

\begin{figure}
\includegraphics[ scale=0.65]{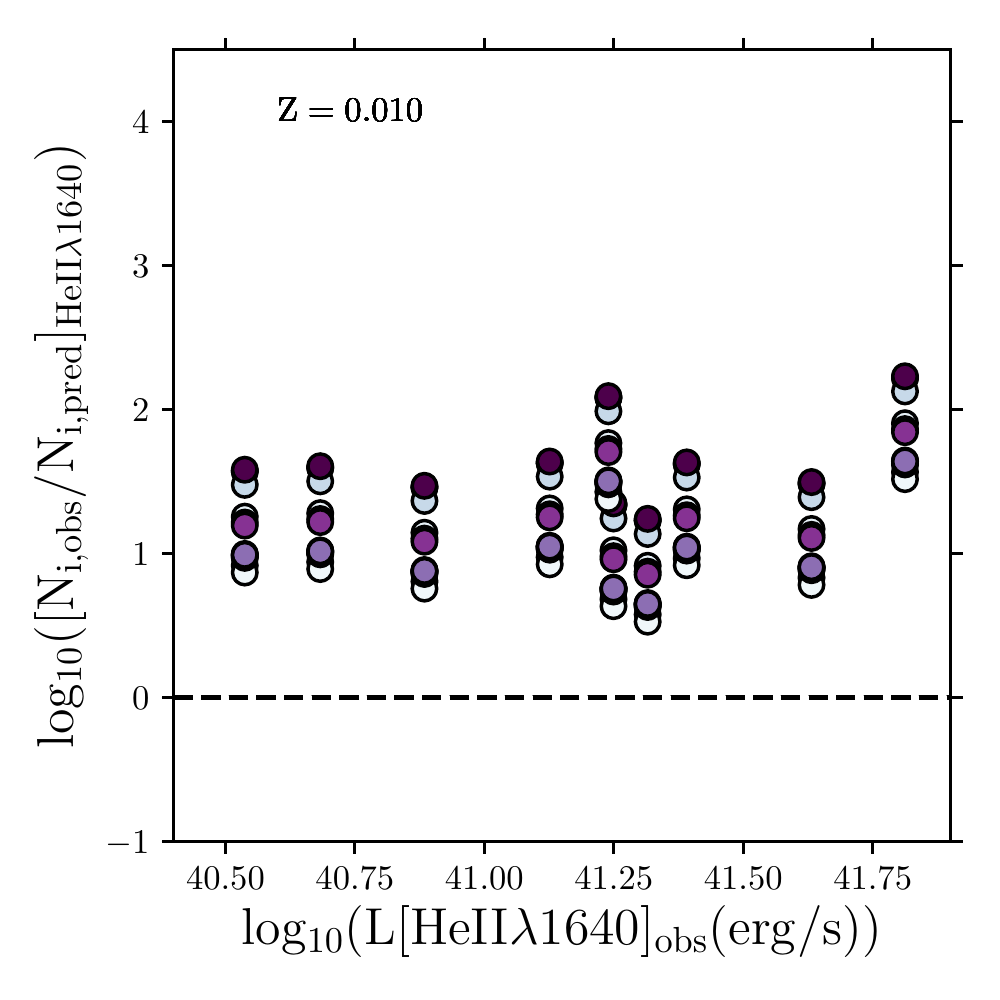}
\includegraphics[ scale=0.65]{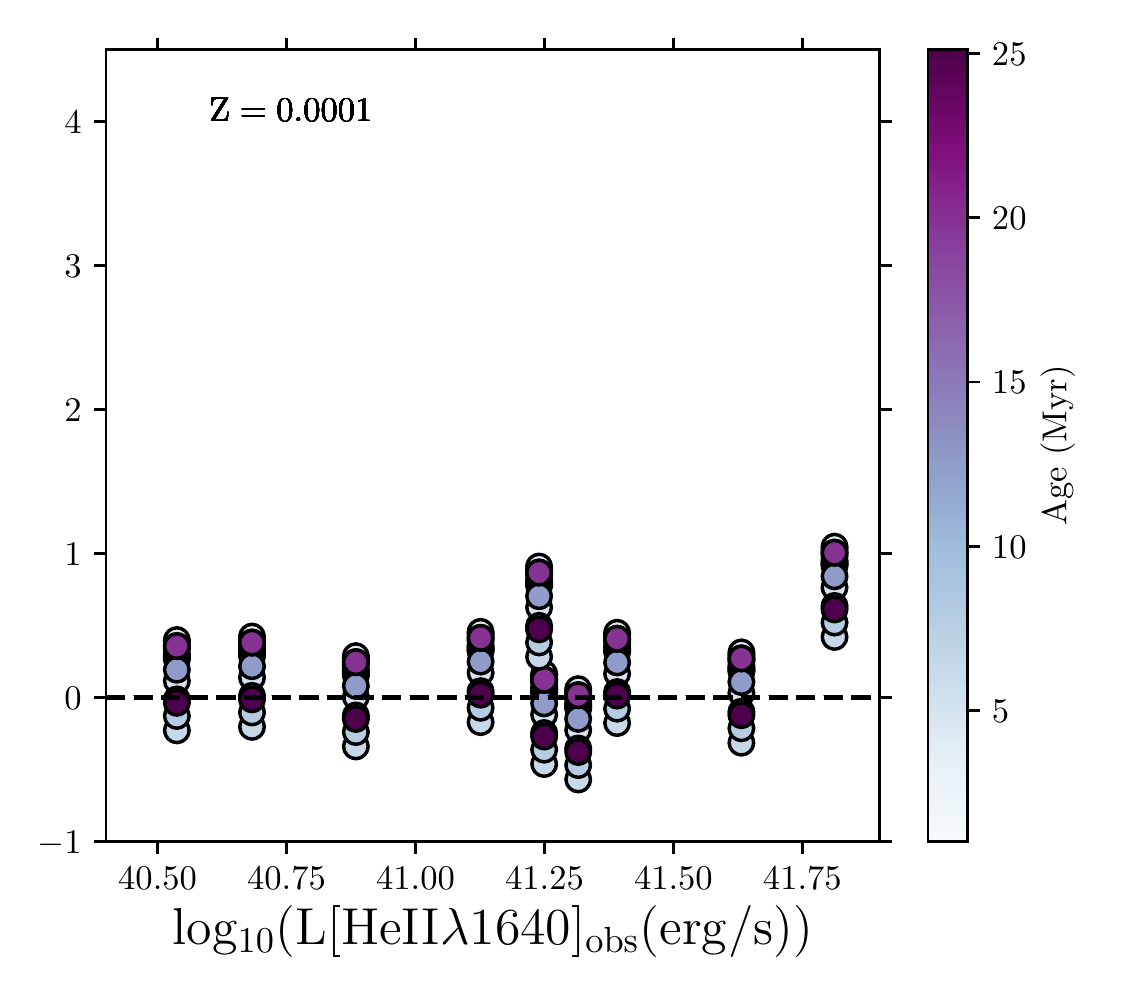}
\caption{The fraction of observed \Hep\ ionising photons compared to \citet{Xiao2018} model expectations as a function of observed \HeII\ luminosity of the MUSE \HeII\ sample. From {\bf left} to {\bf right}, we show the \Hep\ model predictions computed for three metallicities, $Z=0.01, 0.0001$ with $\mathrm{log_{10}}(n_H)=1.0$ and $U_{s}=-1.5$ for different times between $1-20$ Myr from the onset of the star-burst. The dashed horizontal line indicates y=0, where there is no difference between observations and model predictions. 
\label{fig:NHeII_def}
}
\end{figure}

The main discrepancy between model and data arise only once line EWs are compared. As shown by Figure  \ref{fig:ews}, \citet{Xiao2018} binary models are able to reproduce observed \CIII\ EWs but lacks sufficient mechanisms to reproduce the observed \HeII\ EWs. We expect this to be primarily driven by the lack of photons below $\lambda<228$ \AA\ in BPASS models \citep[e.g.,][]{Berg2018}. 
We further develop a simple prescription to investigate the difference in \HeII\ ionising photons between observed data and \citet{Xiao2018} model predictions by normalizing observed \CIII\ luminosities with the models. 
In Figure \ref{fig:NHeII_def} we show the fraction of observed \Hep\ ionising photons compared to the predictions from the models.
Only extreme sub-solar metallicities ($\sim1/200$th) are able to accurately predict the observed \Hep\ ionising photons, which is strongly in contrast with predictions from line-ratio diagnostics.

\section{Conclusions \& Future directions}

% Driven by the advancement in instrumentation in current 8-10m class telescopes, we are able to push our current generation of telescopes to their limits to open up a never observed observational spaces. 
Here, we have used deep optical spectroscopy from MUSE to obtain a sample of \HeII\ detections at $z\sim2-4$ to study their ISM conditions using state-of-the-art stellar-population/photo-ionization models. 
Using rest-UV emission-line ratio diagnostics we show that our galaxies could mostly be explained by Z$\sim0.05-1.0$ \zsol\ photo-ionisation models, but, we show that even BPASS binary models lack sufficient ionising photons to re-produce observed \HeII\ EWs. Using a simple prescription, we show that our observed \HeII\ luminosities can only be explained using extreme sub-solar metallicities ($\sim1/200$th). 
Such low metallicities are in contradiction with our line-ratio diagnostics and stellar populations models can suffer large uncertainties due to lack of empirical calibrations in this regime. 
It is possible that extra contribution from X-Ray binaries, sub-dominant AGN, or effects related to stellar rotations at high metallicities can supply the missing ionising photons. 
Alternatively, if star-forming galaxies at $z\sim2-4$ have a top-heavy initial-mass-function \citep[see][]{Nanayakkara2017}, the extra O/B type stars will contribute to higher levels of ionising photons, which could increase the \Hep\ photon budget. 

Future deep surveys such as the MUSE extreme deep field survey, a single 160 hour pointing by MUSE, will provide extremely high signal-to-noise rest-UV spectra at $z=2-4$ to perform spectro-photometric analysis by simultaneous combination of nebular emission features with weaker ISM and photospheric emission and absorption features. Thus, we will be able to constrain stellar population properties to finer detail within this epoch and make predictions for future surveys by the \emph{James Webb Space Telescope}. Given that individual detections of pop-III stars will be unlikely until proposed future space telescopes such as LUVOIR, we should push the current instruments to their maximum potential to constrain the stellar population properties of galaxies leading to the buildup of the peak of the cosmic star-formation rate density.

\begin{multicols}{2}
\bibliographystyle{yahapj}
\begingroup
\setstretch{0.0}
\bibliography{bibliography.bib}

\begin{thebibliography}{}
\providecommand\natexlab[1]{#1}
\providecommand\JournalTitle[1]{#1}

\bibitem[{{Bacon} {et~al.}(2010){Bacon}, {Accardo}, {Adjali}, {Anwand},
  {Bauer}, {Biswas}, {Blaizot}, {Boudon}, {Brau-Nogue}, {Brinchmann},
  {Caillier}, {Capoani}, {Carollo}, {Contini}, {Couderc}, {Daguis{\'e}},
  {Deiries}, {Delabre}, {Dreizler}, {Dubois}, {Dupieux}, {Dupuy}, {Emsellem},
  {Fechner}, {Fleischmann}, {Fran{\c c}ois}, {Gallou}, {Gharsa}, {Glindemann},
  {Gojak}, {Guiderdoni}, {Hansali}, {Hahn}, {Jarno}, {Kelz}, {Koehler},
  {Kosmalski}, {Laurent}, {Le Floch}, {Lilly}, {Lizon}, {Loupias}, {Manescau},
  {Monstein}, {Nicklas}, {Olaya}, {Pares}, {Pasquini}, {P{\'e}contal-Rousset},
  {Pell{\'o}}, {Petit}, {Popow}, {Reiss}, {Remillieux}, {Renault}, {Roth},
  {Rupprecht}, {Serre}, {Schaye}, {Soucail}, {Steinmetz}, {Streicher}, {Stuik},
  {Valentin}, {Vernet}, {Weilbacher}, {Wisotzki}, \& {Yerle}}]{Bacon2010}
{Bacon}, R., {Accardo}, M., {Adjali}, L., {et~al.} 2010,
  \href{http://dx.doi.org/10.1117/12.856027}{in \procspie, Vol. 7735,
  Ground-based and Airborne Instrumentation for Astronomy III}, 773508

\bibitem[{{Bacon} {et~al.}(2015){Bacon}, {Brinchmann}, {Richard}, {Contini},
  {Drake}, {Franx}, {Tacchella}, {Vernet}, {Wisotzki}, {Blaizot}, {Bouch{\'e}},
  {Bouwens}, {Cantalupo}, {Carollo}, {Carton}, {Caruana}, {Cl{\'e}ment},
  {Dreizler}, {Epinat}, {Guiderdoni}, {Herenz}, {Husser}, {Kamann}, {Kerutt},
  {Kollatschny}, {Krajnovic}, {Lilly}, {Martinsson}, {Michel-Dansac},
  {Patricio}, {Schaye}, {Shirazi}, {Soto}, {Soucail}, {Steinmetz}, {Urrutia},
  {Weilbacher}, \& {de Zeeuw}}]{Bacon2015}
{Bacon}, R., {Brinchmann}, J., {Richard}, J., {et~al.} 2015,
  \href{http://dx.doi.org/10.1051/0004-6361/201425419}{\JournalTitle{\aap},
  575, A75}

\bibitem[{{Bacon} {et~al.}(2017){Bacon}, {Conseil}, {Mary}, {Brinchmann},
  {Shepherd}, {Akhlaghi}, {Weilbacher}, {Piqueras}, {Wisotzki}, {Lagattuta},
  {Epinat}, {Guerou}, {Inami}, {Cantalupo}, {Courbot}, {Contini}, {Richard},
  {Maseda}, {Bouwens}, {Bouch{\'e}}, {Kollatschny}, {Schaye}, {Marino},
  {Pello}, {Herenz}, {Guiderdoni}, \& {Carollo}}]{Bacon2017}
{Bacon}, R., {Conseil}, S., {Mary}, D., {et~al.} 2017,
  \href{http://dx.doi.org/10.1051/0004-6361/201730833}{\JournalTitle{\aap},
  608, A1}

\bibitem[{{Berg} {et~al.}(2018){Berg}, {Erb}, {Auger}, {Pettini}, \&
  {Brammer}}]{Berg2018}
{Berg}, D.~A., {Erb}, D.~K., {Auger}, M.~W., {Pettini}, M., \& {Brammer}, G.~B.
  2018, \JournalTitle{ArXiv e-prints},
  \href{http://arxiv.org/abs/1803.02340}{{\sffamily arXiv:1803.02340}}

\bibitem[{{Binette} {et~al.}(1994){Binette}, {Magris}, {Stasi{\'n}ska}, \&
  {Bruzual}}]{Binette1994}
{Binette}, L., {Magris}, C.~G., {Stasi{\'n}ska}, G., \& {Bruzual}, A.~G. 1994,
  \JournalTitle{\aap}, 292, 13

\bibitem[{{Casares} {et~al.}(2017){Casares}, {Jonker}, \&
  {Israelian}}]{Casares2017}
{Casares}, J., {Jonker}, P.~G., \& {Israelian}, G. 2017, \JournalTitle{ArXiv
  e-prints}, \href{http://arxiv.org/abs/1701.07450}{{\sffamily arXiv:1701.07450
  [astro-ph.HE]}}

\bibitem[{{Cassata} {et~al.}(2013){Cassata}, {Le F{\`e}vre}, {Charlot},
  {Contini}, {Cucciati}, {Garilli}, {Zamorani}, {Adami}, {Bardelli}, {Le Brun},
  {Lemaux}, {Maccagni}, {Pollo}, {Pozzetti}, {Tresse}, {Vergani}, {Zanichelli},
  \& {Zucca}}]{Cassata2013}
{Cassata}, P., {Le F{\`e}vre}, O., {Charlot}, S., {et~al.} 2013,
  \href{http://dx.doi.org/10.1051/0004-6361/201220969}{\JournalTitle{\aap},
  556, A68}

\bibitem[{{Eldridge} {et~al.}(2017){Eldridge}, {Stanway}, {Xiao}, {McClelland},
  {Taylor}, {Ng}, {Greis}, \& {Bray}}]{Eldridge2017}
{Eldridge}, J.~J., {Stanway}, E.~R., {Xiao}, L., {et~al.} 2017,
  \href{http://dx.doi.org/10.1017/pasa.2017.51}{\JournalTitle{PASA}, 34, e058}

\bibitem[{{Epinat} {et~al.}(2018){Epinat}, {Contini}, {Finley}, {Boogaard},
  {Gu{\'e}rou}, {Brinchmann}, {Carton}, {Michel-Dansac}, {Bacon}, {Cantalupo},
  {Carollo}, {Hamer}, {Kollatschny}, {Krajnovi{\'c}}, {Marino}, {Richard},
  {Soucail}, {Weilbacher}, \& {Wisotzki}}]{Epinat2018}
{Epinat}, B., {Contini}, T., {Finley}, H., {et~al.} 2018,
  \href{http://dx.doi.org/10.1051/0004-6361/201731877}{\JournalTitle{\aap},
  609, A40}

\bibitem[{{G{\"o}tberg} {et~al.}(2017){G{\"o}tberg}, {de Mink}, \&
  {Groh}}]{Gotberg2017}
{G{\"o}tberg}, Y., {de Mink}, S.~E., \& {Groh}, J.~H. 2017,
  \href{http://dx.doi.org/10.1051/0004-6361/201730472}{\JournalTitle{\aap},
  608, A11}

\bibitem[{{Gutkin} {et~al.}(2016){Gutkin}, {Charlot}, \&
  {Bruzual}}]{Gutkin2016}
{Gutkin}, J., {Charlot}, S., \& {Bruzual}, G. 2016,
  \href{http://dx.doi.org/10.1093/mnras/stw1716}{\JournalTitle{\mnras}, 462,
  1757}

\bibitem[{{Heger} \& {Woosley}(2002)}]{Heger2002}
{Heger}, A., \& {Woosley}, S.~E. 2002,
  \href{http://dx.doi.org/10.1086/338487}{\JournalTitle{\apj}, 567, 532}

\bibitem[{{Inami} {et~al.}(2017){Inami}, {Bacon}, {Brinchmann}, {Richard},
  {Contini}, {Conseil}, {Hamer}, {Akhlaghi}, {Bouch{\'e}}, {Cl{\'e}ment},
  {Desprez}, {Drake}, {Hashimoto}, {Leclercq}, {Maseda}, {Michel-Dansac},
  {Paalvast}, {Tresse}, {Ventou}, {Kollatschny}, {Boogaard}, {Finley},
  {Marino}, {Schaye}, \& {Wisotzki}}]{Inami2017}
{Inami}, H., {Bacon}, R., {Brinchmann}, J., {et~al.} 2017,
  \href{http://dx.doi.org/10.1051/0004-6361/201731195}{\JournalTitle{\aap},
  608, A2}

\bibitem[{{Inoue} {et~al.}(2011){Inoue}, {Kousai}, {Iwata}, {Matsuda},
  {Nakamura}, {Horie}, {Hayashino}, {Tapken}, {Akiyama}, {Noll}, {Yamada},
  {Burgarella}, \& {Nakamura}}]{Inoue2011}
{Inoue}, A.~K., {Kousai}, K., {Iwata}, I., {et~al.} 2011,
  \href{http://dx.doi.org/10.1111/j.1365-2966.2010.17851.x}{\JournalTitle{\mnras},
  411, 2336}

\bibitem[{{Izotov} {et~al.}(2012){Izotov}, {Thuan}, \& {Privon}}]{Izotov2012}
{Izotov}, Y.~I., {Thuan}, T.~X., \& {Privon}, G. 2012,
  \href{http://dx.doi.org/10.1111/j.1365-2966.2012.22051.x}{\JournalTitle{\mnras},
  427, 1229}

\bibitem[{{Kacprzak} {et~al.}(2016){Kacprzak}, {van de Voort}, {Glazebrook},
  {Tran}, {Yuan}, {Nanayakkara}, {Allen}, {Alcorn}, {Cowley}, {Labb{\'e}},
  {Spitler}, {Straatman}, \& {Tomczak}}]{Kacprzak2016}
{Kacprzak}, G.~G., {van de Voort}, F., {Glazebrook}, K., {et~al.} 2016,
  \href{http://dx.doi.org/10.3847/2041-8205/826/1/L11}{\JournalTitle{\apjl},
  826, L11}

\bibitem[{{Kewley} {et~al.}(2016){Kewley}, {Yuan}, {Nanayakkara}, {Kacprzak},
  {Tran}, {Glazebrook}, {Spitler}, {Cowley}, {Dopita}, {Straatman},
  {Labb{\'e}}, \& {Tomczak}}]{Kewley2016}
{Kewley}, L.~J., {Yuan}, T., {Nanayakkara}, T., {et~al.} 2016,
  \href{http://dx.doi.org/10.3847/0004-637X/819/2/100}{\JournalTitle{\apj},
  819, 100}

\bibitem[{{Madau} \& {Dickinson}(2014)}]{Madau2014}
{Madau}, P., \& {Dickinson}, M. 2014,
  \href{http://dx.doi.org/10.1146/annurev-astro-081811-125615}{\JournalTitle{\araa},
  52, 415}

\bibitem[{{Marino} {et~al.}(2018){Marino}, {Cantalupo}, {Lilly}, {Gallego},
  {Straka}, {Borisova}, {Pezzulli}, {Bacon}, {Brinchmann}, {Carollo},
  {Caruana}, {Conseil}, {Contini}, {Diener}, {Finley}, {Inami}, {Leclercq},
  {Muzahid}, {Richard}, {Schaye}, {Wendt}, \& {Wisotzki}}]{Marino2018}
{Marino}, R.~A., {Cantalupo}, S., {Lilly}, S.~J., {et~al.} 2018,
  \href{http://dx.doi.org/10.3847/1538-4357/aab6aa}{\JournalTitle{\apj}, 859,
  53}

\bibitem[{{Matthee} {et~al.}(2017){Matthee}, {Sobral}, {Boone},
  {R{\"o}ttgering}, {Schaerer}, {Girard}, {Pallottini}, {Vallini}, {Ferrara},
  {Darvish}, \& {Mobasher}}]{Matthee2017}
{Matthee}, J., {Sobral}, D., {Boone}, F., {et~al.} 2017,
  \href{http://dx.doi.org/10.3847/1538-4357/aa9931}{\JournalTitle{\apj}, 851,
  145}

\bibitem[{{Naidu} {et~al.}(2017){Naidu}, {Oesch}, {Reddy}, {Holden}, {Steidel},
  {Montes}, {Atek}, {Bouwens}, {Carollo}, {Cibinel}, {Illingworth},
  {Labb{\'e}}, {Magee}, {Morselli}, {Nelson}, {van Dokkum}, \&
  {Wilkins}}]{Naidu2017}
{Naidu}, R.~P., {Oesch}, P.~A., {Reddy}, N., {et~al.} 2017,
  \href{http://dx.doi.org/10.3847/1538-4357/aa8863}{\JournalTitle{\apj}, 847,
  12}

\bibitem[{{Nanayakkara} {et~al.}(2017){Nanayakkara}, {Glazebrook}, {Kacprzak},
  {Yuan}, {Fisher}, {Tran}, {Kewley}, {Spitler}, {Alcorn}, {Cowley}, {Labbe},
  {Straatman}, \& {Tomczak}}]{Nanayakkara2017}
{Nanayakkara}, T., {Glazebrook}, K., {Kacprzak}, G.~G., {et~al.} 2017,
  \href{http://dx.doi.org/10.1093/mnras/stx605}{\JournalTitle{\mnras}, 468,
  3071}

\bibitem[{{Patr{\'{\i}}cio} {et~al.}(2016){Patr{\'{\i}}cio}, {Richard},
  {Verhamme}, {Wisotzki}, {Brinchmann}, {Turner}, {Christensen}, {Weilbacher},
  {Blaizot}, {Bacon}, {Contini}, {Lagattuta}, {Cantalupo}, {Cl{\'e}ment}, \&
  {Soucail}}]{Patricio2017}
{Patr{\'{\i}}cio}, V., {Richard}, J., {Verhamme}, A., {et~al.} 2016,
  \href{http://dx.doi.org/10.1093/mnras/stv2859}{\JournalTitle{\mnras}, 456,
  4191}

\bibitem[{{Senchyna} {et~al.}(2017){Senchyna}, {Stark}, {Vidal-Garc{\'{\i}}a},
  {Chevallard}, {Charlot}, {Mainali}, {Jones}, {Wofford}, {Feltre}, \&
  {Gutkin}}]{Senchyna2017}
{Senchyna}, P., {Stark}, D.~P., {Vidal-Garc{\'{\i}}a}, A., {et~al.} 2017,
  \JournalTitle{ArXiv e-prints},
  \href{http://arxiv.org/abs/1706.00881}{{\sffamily arXiv:1706.00881}}

\bibitem[{{Shibuya} {et~al.}(2017){Shibuya}, {Ouchi}, {Harikane}, {Rauch},
  {Ono}, {Mukae}, {Higuchi}, {Kojima}, {Yuma}, {Lee}, {Furusawa}, {Konno},
  {Martin}, {Shimasaku}, {Taniguchi}, {Kobayashi}, {Kajisawa}, {Nagao}, {Goto},
  {Kashikawa}, {Komiyama}, {Kusakabe}, {Momose}, {Nakajima}, {Tanaka}, \&
  {Wang}}]{Shibuya2017}
{Shibuya}, T., {Ouchi}, M., {Harikane}, Y., {et~al.} 2017, \JournalTitle{ArXiv
  e-prints}, \href{http://arxiv.org/abs/1705.00733}{{\sffamily
  arXiv:1705.00733}}

\bibitem[{{Shirazi} \& {Brinchmann}(2012)}]{Shirazi2012}
{Shirazi}, M., \& {Brinchmann}, J. 2012,
  \href{http://dx.doi.org/10.1111/j.1365-2966.2012.20439.x}{\JournalTitle{\mnras},
  421, 1043}

\bibitem[{{Sobral} {et~al.}(2015){Sobral}, {Matthee}, {Darvish}, {Schaerer},
  {Mobasher}, {R{\"o}ttgering}, {Santos}, \& {Hemmati}}]{Sobral2015}
{Sobral}, D., {Matthee}, J., {Darvish}, B., {et~al.} 2015,
  \href{http://dx.doi.org/10.1088/0004-637X/808/2/139}{\JournalTitle{\apj},
  808, 139}

\bibitem[{{Sobral} {et~al.}(2017){Sobral}, {Matthee}, {Brammer}, {Ferrara},
  {Alegre}, {Rottgering}, {Schaerer}, {Mobasher}, \& {Darvish}}]{Sobral2017}
{Sobral}, D., {Matthee}, J., {Brammer}, G., {et~al.} 2017, \JournalTitle{ArXiv
  e-prints}, \href{http://arxiv.org/abs/1710.08422}{{\sffamily
  arXiv:1710.08422}}

\bibitem[{{Steidel} {et~al.}(2016){Steidel}, {Strom}, {Pettini}, {Rudie},
  {Reddy}, \& {Trainor}}]{Steidel2016}
{Steidel}, C.~C., {Strom}, A.~L., {Pettini}, M., {et~al.} 2016,
  \href{http://dx.doi.org/10.3847/0004-637X/826/2/159}{\JournalTitle{\apj},
  826, 159}

\bibitem[{{Strom} {et~al.}(2017){Strom}, {Steidel}, {Rudie}, {Trainor},
  {Pettini}, \& {Reddy}}]{Strom2017}
{Strom}, A.~L., {Steidel}, C.~C., {Rudie}, G.~C., {et~al.} 2017,
  \href{http://dx.doi.org/10.3847/1538-4357/836/2/164}{\JournalTitle{\apj},
  836, 164}

\bibitem[{{Xiao} {et~al.}(2018){Xiao}, {Stanway}, \& {Eldridge}}]{Xiao2018}
{Xiao}, L., {Stanway}, E., \& {Eldridge}, J.~J. 2018, \JournalTitle{ArXiv
  e-prints}, \href{http://arxiv.org/abs/1801.07068}{{\sffamily
  arXiv:1801.07068}}

\end{thebibliography}
\endgroup
\end{multicols}

\end{document}